\documentclass[letterpaper,11pt]{article}
\usepackage{jheppub}
\pdfoutput=1

\usepackage{amsfonts}
\usepackage{amsmath}
\usepackage{amssymb}
\usepackage{comment}
\usepackage{float}
\usepackage{graphicx} 

\graphicspath{{./images/}}

\title{Allowable Complex Black Holes in the 
Euclidean Gravitational Path Integral}
\author{Vineeth Krishna and Finn Larsen}
\affiliation{Leinweber Institute for Theoretical Physics, University of Michigan, Ann Arbor, MI, USA 48109}
\emailAdd{vkt@umich.edu}
\emailAdd{larsenf@umich.edu}

\abstract{
The Euclidean Gravitational Path Integral has proven remarkably effective in the quantum regime of black hole physics. In this work, we examine the applicability of the Kontsevich-Segal-Witten (KSW) criterion for admissible complex metrics in the context of the Euclidean Gravitational Path Integral. We find that, for the super-conformal index of ${\cal N}=4$ SYM with unequal angular momenta, the black hole saddle points violate the KSW criterion precisely where the statistical description of the index breaks down. The corresponding critical point coincides with a phase transition into two-component ``grey galaxy'' configurations. 
}

\begin{document}

\maketitle
\flushbottom

\section{Introduction}

The Euclidean Gravitational Path Integral (EGPI), formally defined in the seminal work of Gibbons and Hawking \cite{Gibbons:1976ue}, has historically produced results well beyond what one could expect from a semi-classical approximation. For many years, the EGPI was primarily utilized to derive the thermodynamics of black holes \cite{Hawking:1982dh,York:1986it,Whiting:1988qr,Brown:1992br}, interpreting the path integral as a partition function in the canonical ensemble where the saddle point approximation $Z \approx e^{-I_{E}}$ yields the Bekenstein-Hawking entropy. However, in recent years, this tool has experienced a resurgence \cite{Sen:2007qy,Sen:2008vm,Sen:2009vz,Banerjee:2010qc,Sen:2012cj,Iliesiu:2020qvm,Heydeman:2020hhw,Iliesiu:2022kny,Iliesiu:2022onk,Boruch:2022tno,Boruch:2023gfn,Cassani:2024kjn,Boruch:2025qdq,Boruch:2025biv,Cassani:2025iix,Boruch:2025sie}, proving remarkably successful in decoding the nature of quantum corrections in near-extremal and near-BPS black holes previously thought inaccessible.

A significant development occurred with the realization that the path integral must include contributions from topologies beyond the standard black hole saddle. Major progress towards resolving the Black Hole Information Paradox was achieved by considering the contribution of so-called replica wormhole saddles \cite{Almheiri:2019hni, Penington:2019kki}, which are essential for recovering the Page curve required by unitarity. Similarly, signatures of the chaotic nature of black holes — specifically the ramp and plateau in the spectral form factor — can be recovered from the double cone geometry \cite{Saad:2018bqo,Chen:2023hra}. Yet, this success brings with it a fundamental challenge. With the understanding that novel saddle points like wormholes lead to essential new physics, the uncertainty surrounding the precise rules governing the gravitational path integral becomes untenable. In particular, it is not immediately clear which saddle points should be included in the summation, and which should be discarded as unphysical.

The AdS/CFT correspondence \cite{Maldacena:1997re,Witten:1998qj} offers a rich testing ground for us
to constrain these rules. By leveraging the holographic duality, one can demand that the behavior of the bulk path integral matches the computation in the dual field theory. However, since the correspondence maps a weakly interacting gravitational theory in the bulk to a strongly interacting boundary Conformal Field Theory (CFT), the quantity one studies must be computable at large values of the CFT coupling constant. Superconformal field theories furnish one such protected quantity — the superconformal index (SCI) \cite{Romelsberger:2005eg,Kinney:2005ej}. As a partition function over BPS states, the SCI is defined as a function of chemical potentials (fugacities) associated with the Cartan generators that commute with the preserved supercharge. From the general representation theory of states in superconformal theories, we can robustly identify the region in the space of these chemical potentials where the index convergences. The boundary of this region is due to a physical phase transition that can be diagnosed 
by the index \cite{Choi:2025lck,Deddo:2025jrg}.  

Using holography, we can compute the index purely in the bulk by choosing appropriate boundary conditions determined by the chemical potentials. The corresponding bulk quantity is termed the \textit{gravitational index}. The dominant saddle points contributing to the gravitational index have been identified as non-extremal analytic continuations of BPS black hole solutions \cite{Cabo-Bizet:2018ehj} . Crucially, such an analytic continuation necessarily makes the spacetime metric complex. The primary goal of this work is to test a specific proposal for the inclusion of such complex metrics: the criterion proposed by Kontsevich and Segal, and developed further by Witten (KSW) \cite{Kontsevich:2021dmb,Witten:2021nzp}. We show that the KSW criteria are satisfied precisely if and only if the index is well-defined. 

The physical regions we determine in several ways are nontrivial for $\mathcal{N}=4$ Supersymmetric Yang-Mills when the angular velocities are unequal, but electric charges are the same\footnote{The KSW criteria for unequal angular momenta was previously studied in \cite{BenettiGenolini:2025jwe}. However, they concluded that the KSW conditions are insufficient, because of an error in the calculation of eigenvalues.}. In the analogous computation for complex Euclidean $AdS_4$ black hole saddles with two pairs of equal charges, the KSW conditions are insufficient, they are too weak to rule out complex saddles in regions of chemical potential space where the boundary SCI is known to diverge \cite{Jones:2025gno}. We expect that the KSW conditions will similarly be insufficient for AdS$_5$ black holes with unequal electric charges. 

This paper is organized as follows: In Section \ref{sec:EGPI}, we introduce the Euclidean Gravitational Path Integral and the Superconformal Index, identifying the precise region of convergence for the SCI in the space of complex chemical potentials. In Section \ref{sec:KSW} we develop a practical method for implementing the KSW criterion to determine admissible complex metrics. In Section \ref{sec:uneqj} we introduce the Lorentzian and Euclidean versions of the AdS$_5$ black holes with two independent angular momenta and a single electric charge. We identify the solutions that preserve supersymmetry and the subset of those that satisfy the KSW conditions. Those are the allowed complex saddle points that contribute of the gravitational index.
Finally, in Section \ref{sec:discussion}, we compare to other recent work and discuss open problems. 

\bigskip
\noindent \textbf{Note:} During the completion of this work, the authors of \cite{BenettiGenolini:2025jwe} posted a corrected version of their computation in \cite{BenettiGenolini:2026raa} which has a significant overlap with the work in this paper.

\section{The Euclidean Gravitational Path Integral}\label{sec:EGPI}

In this section we review the concept of the Euclidean Gravitational Path Integral (EGPI). We define the Superconformal Index in $\mathcal{N}=4$ supersymmetric Yang-Mills theory and obtain the region of convergence of the Index in the space of complex chemical potentials. In the context of holography, the Superconformal Index (SCI) of the boundary CFT can be interpreted as an EGPI evaluated on bulk geometries that satisfy specific boundary conditions. We discuss the Kontsevich-Segal-Witten (KSW) criteria \cite{Kontsevich:2021dmb,Witten:2021nzp} which determines the applicable complex saddle points of the EGPI. 

\subsection{Euclidean Gravitational Path Integral and Saddle Points}

The Euclidean Gravitational Path Integral (EGPI), originally proposed by Gibbons and Hawking \cite{Gibbons:1976ue}, computes the partition function of a quantum theory of gravity. Formally, it is defined as a path integral over the space of metrics and matter fields, weighted by the Euclidean gravitational action:
\begin{equation}\label{EGPI}
    \mathcal{Z} = \int \mathcal{D}g_{\mu\nu} \mathcal{D}\Psi \, e^{-I_E[g_{\mu\nu},\Psi]}~,
\end{equation}
where $I_E[g_{\mu\nu},\Psi]$ denotes the Euclidean action, $g_{\mu\nu}$ is the spacetime metric, and $\Psi$ represents the collection of all other dynamic fields. All fields are subject to boundary conditions.

Two primary approaches exist to evaluating \eqref{EGPI}. The first approximates the path integral via a sum over saddle points. The second leverages the AdS/CFT correspondence to equate \eqref{EGPI} with a trace over the Hilbert space of the boundary CFT. Our objective is to substantiate the validity of the first approach by verifying that properties observed in the boundary CFT are reproduced by the bulk saddle point analysis.

Although a rigorous mathematical definition of the path integral over the space of metrics remains elusive, we assume the existence of a suitable definition. In the semiclassical limit ($G_N \rightarrow 0$), the EGPI \eqref{EGPI} is approximated by the sum of contributions from dominant saddle points
\begin{equation}
    \mathcal{Z} \approx e^{-I_1} + e^{-I_2} + \ldots
\end{equation}
Here, $I_i$ represents the on-shell Euclidean action evaluated at the $i$-th saddle point—solutions to the equations of motion derived from $I_E[g_{\mu\nu},\Psi]$ that satisfy the imposed boundary conditions. The leading contribution to the EGPI arises from the saddle point minimizing the real part of the on-shell action. While this interpretation is straightforward when the on-shell actions $I_i$ are real, there is no \textit{a priori} reason to restrict the integration contour to the space of real metrics. Indeed complex saddle points play a crucial role in many EGPI computations.

\subsection{The Superconformal Index}\label{sec:SCI}
In the AdS/CFT correspondance, physical variables in the weakly coupled gravitational theory are mapped to observables in a CFT that is intractible, because it is strongly coupled. The Superconformal Index (SCI) allows for meaningful comparison, because it can be computed reliably at any coupling. We focus on the duality between Type IIB String Theory on asymptotically $AdS_5\times S^5$ and $\mathcal{N}=4$ Supersymmetric Yang-Mills (SYM) on $S^3 \times  \mathbb{R}$. In this case the SCI is defined as the trace over the CFT Hilbert space
\begin{equation}
    \mathcal{I}(\Omega_a^\prime, \Omega_b^\prime, \Phi_1^\prime, \Phi_2^\prime, \Phi_3^\prime) = \mathrm{Tr} \left[e^{-\beta\{\mathcal Q,\mathcal Q^\dagger\}} e^{-\Omega^\prime_a J_a -\Omega^\prime_b J_b -\Phi^\prime_1 Q_1-\Phi^\prime_2 Q_2-\Phi^\prime_3 Q_3 }\right]~. \label{indexdefn}
\end{equation}
The familiar insertion of $(-)^F$ is implemented by imposing the following constraint on the arguments of the index:
\begin{equation}\label{indexcond}
    \Omega_a^\prime + \Omega^\prime_b - \Phi_1^\prime -\Phi_2^\prime-\Phi_3^\prime = 2\pi i (2\mathbb Z + 1)~.
\end{equation}
This condition ensures that states related by the action of the supercharge $\mathcal{Q}$ cancel in the trace. Consequently, all non-vanishing contributions arise from BPS states annihilated by $\mathcal{Q}$. Therefore, the index does not depend on the variable $\beta$. In fact, due to the additional constraint \eqref{indexcond} the SCI \eqref{indexdefn} actually depends only on four (complex) chemical potentials.

The superconformal index \eqref{indexdefn} can also be written as
\begin{align}
    \mathcal{I} = \mathrm{Tr} \left[e^{-\beta(E-\Omega_aJ_a-\Omega_bJ_b-\Phi_1Q_1-\Phi_2Q_2-\Phi_3Q_3)}\right]~, \label{indtrace}
\end{align}
with the constraint \eqref{indexcond} taking the form
\begin{equation}
    \beta(1+\Omega_a + \Omega_b - \Phi_1 -\Phi_2 - \Phi_3) = 2\pi i (2\mathbb Z+1)~.
\end{equation}
We can recover the earlier definition \eqref{indexdefn} through the identifications
\begin{align}
    \Omega_{a,b}^\prime &= \beta(1-\Omega_{a,b})~,\label{omegaiden}\\
    \Phi_{1,2,3}^\prime &= \beta(1-\Phi_{1,2,3})~,\label{phiiden}\\
    \{\mathcal{Q},\mathcal{Q}^\dagger\} &= E-J_a-J_b-Q_1-Q_2-Q_3~.
\end{align}

To evaluate the trace \eqref{indtrace} using a path integral, we consider the $CFT$ on $S^3\times \mathbb R$ and Wick rotate it to Euclidean signature by taking the time coordinate in the Lorentzian $CFT$ $t=-it_E$, where $t_E$ is the Euclidean time in the Wick rotated $CFT$. The fugacities $(\beta,\Omega_a,\Omega_b)$ are introduced by imposing periodic identifications on the $CFT$ coordinates,
\begin{align}
    (\tau, \phi_a, \phi_b) \sim (\tau+\beta, \phi_a-i\beta\Omega_a, \phi_b-i\beta\Omega_b)\label{twistident}
\end{align}
and the chemical potentials $\Phi_{1,2,3}$ are fixed by fixing the holonomy of the gauge fields $A_i$ around the thermal circle.

In section \ref{sec:uneqj}, we restrict our analysis to the special case of the superconformal index where the three chemical potentials $\Phi_{1,2,3}^\prime$ are taken to be equal and denoted $\Phi^\prime$. In this sub-sector, the superconformal index is
\begin{align}
    \mathcal{I}(\Omega_a^\prime,\Omega_b^\prime,\Phi^\prime) = \mathrm{Tr}\left[e^{-\beta\{\mathcal{Q},\mathcal{Q}^\dagger\}} e^{-\Omega_a^\prime J_a - \Omega_b^\prime J_b - 3\Phi^\prime Q}\right]~,
\end{align}
where $Q=\frac{Q_1+Q_2+Q_3}{3}$ and the constraint on the chemical potentials becomes
\begin{equation}
    \Omega_a^\prime + \Omega^\prime_b - 3\Phi^\prime= 2\pi i (2\mathbb Z + 1)~~.
\end{equation}

\subsection{Convergence Conditions on the Index}\label{sec:converge}

The partition function \eqref{indexdefn} defined as a trace over the entire Hilbert space of a CFT is not well-defined for chemical potentials in the entire complex plane. The Hilbert space of a CFT decomposes into modules, where each module contains a primary operator and all of its descendants. For a primary operator $\mathcal{O}$ in $\mathcal{N}=4$ SYM with the charges $(J_a,J_b,Q_1,Q_2,Q_3)$, the tower of descendants have scaling dimensions $E$ that are not bounded from above. Therefore, convergence of the index \eqref{indtrace} requires a restriction on the allowed values of $\beta$: 
\begin{equation}\label{betabound}
    \operatorname{Re}(\beta)>0 \,. 
\end{equation}
This is needed to ensure that operators
with large scaling dimension are suppressed in the trace, rather than enhanced. Although we are ultimately interested in supersymmetric states, for which the coefficient of $\beta$ in \eqref{indexdefn} vanishes, the cancellation of contributions from every pair of bosonic and fermionic non-supersymmetric states is properly defined only when the regulator $\beta$ satisfies the condition \eqref{betabound}.

The existence of descendant operators that act as derivatives on primaries is guaranteed by the representation theory of the $\mathcal{N}=4$ superconformal algebra. There are two such derivative operators, $\partial_{+\pm}$ that commute with the preserved supercharges
$\mathcal{Q}$. Their quantum numbers are
\begin{equation}
    (E,J_a,J_b,Q_1,Q_2,Q_3) = (1,1,0,0,0,0) \quad \text{and} \quad (1,0,1,0,0,0)\,.
\end{equation}
The action of these supersymmetric derivatives on a given primary produces a new supersymmetric state with either $J_a$ or $J_b$ increased by a unit. An arbitrary number of these two supersymmetric derivatives can act on a given supersymmetric primary. Therefore, the contributions to the superconformal index coming from the tower of supersymmetric descendant operators of supersymmetric primaries remain finite only when
\begin{equation}\label{omegabound}
    \operatorname{Re}(\Omega_{a,b}^\prime)>0 \,.
\end{equation}
An analogous condition for the chemical potentials $\Phi_{1,2,3}^\prime$ follows by considering the contributions coming from a tower of supersymmetric primaries of the form
\begin{equation}\label{chiralprim}
    {\rm Tr} (X^n) \qquad \text{ for } n\geq 2~. 
\end{equation}
Here $X$ is one of the three supersymmetric scalars that carry the quantum numbers 
\begin{equation}
    (E,J_a,J_b,Q_1,Q_2,Q_3) = (1,0,0,1,0,0)~, \quad (1,0,0,0,1,0)~< \quad \text{and} \quad (1,0,0,0,0,1)\,.
\end{equation}
The supersymmetric primaries of the form \eqref{chiralprim} can be constructed for arbitrary values of $n\geq2$, and for any of the three scalars. The contributions coming from such infinite towers of operators remain finite only when
\begin{equation}\label{phibound}
    {\rm Re}(\Phi_{1,2,3}^\prime)>0~.
\end{equation}
The inequalities \eqref{betabound}, \eqref{omegabound} and \eqref{phibound} together bound the region of chemical potential space where the superconformal index of the $\mathcal{N}=4$ Supersymmetric Yang-Mills theory is expected to converge. The goal of this paper is to realize these boundaries through a bulk computation that identifies instabilities of the saddle points that contribute to the superconformal index. 

\subsection{Allowable Complex Euclidean Saddles}\label{sec:KSW}

The superconformal index defined in section \ref{sec:SCI} as a trace over the Hilbert space of the CFT can be interpreted as an Euclidean Gravitational Path Integral in the bulk theory using the AdS/CFT correspondence. The bulk EGPI which computes the index is called the \textit{gravitational index}. 
The asymptotic boundary conditions for the gravitational index are given by the identifications \eqref{twistident} and the corresponding holonomies in terms of $\Phi_{1,2,3}^\prime$. The constraint on the chemical potentials \eqref{indexcond} is implemented in the bulk gravitational path integral by imposing \textit{periodic} boundary conditions on the fermions. The saddle points for the gravitational index necessarily have a complex metric and are obtained as analytic continuation of the real extremal supersymmetric black hole solutions to Euclidean signature at finite temperature \cite{Cabo-Bizet:2018ehj}. The free energy of these complex saddle point solutions take the standard form \cite{Hosseini:2017mds} required to reproduce the entropy of the corresponding BPS black holes.  

To determine which complex metrics are physically admissible in the gravitational path integral, we use the KSW criterion developed in \cite{Kontsevich:2021dmb,Witten:2021nzp}. An allowable complex Euclidean metric $g_{ij}$ must satisfy the following pointwise condition on the Euclidean manifold
\begin{equation}\label{KSW1}
    \mathrm{Re} \left(\sqrt{\det g} \,g^{i_1j_1}\ldots g^{i_pj_p} F_{i_1\ldots i_p} F_{j_1\ldots j_p}\right) > 0~,
\end{equation}
for any real, non-zero $p$-form $F$. The KSW conditions \eqref{KSW1} consider fluctuations of auxiliary fields around a background metric, but they do not take back reaction into account, and they do not consider fluctuations of the metric itself. Thus the KSW criteria are relatively mild, they are far from the most stringent criteria that one might impose on the admissibility of a saddle point. 

Direct application of the KSW conditions \eqref{KSW1} is often impractical for $p \geq 2$, but the condition can be reformulated in an equivalent form that is more convenient. In a coordinate basis where the metric is diagonal $g_{ij} = \lambda_i \delta_{ij}$, the metric is allowable if
\begin{equation}\label{KSW2}
    \sum_i |\arg \lambda_i| < \pi~.
\end{equation}
Note that the parameters $\lambda_i$ are not eigenvalues of the metric in a conventional sense. The metric is symmetric but, because it is complex, it is not Hermitean. Additionally, the metric transforms as a tensor $g \rightarrow J^T g J$, where $J$ is the Jacobian, so the $\lambda_i$'s are coordinate-dependent. 

To utilize \eqref{KSW2}, we must compute the diagonal entries $\lambda_i$ in the basis where $g_{ij}$ diagonalizes\footnote{See also Appendix B of \cite{Chryssanthacopoulos}.}. We divide our computations into two steps. First, we enforce the $p=0, \text{ and } p=1$ conditions
\begin{align}
    &\mathrm{Re} \left(\sqrt{\det g}\right) > 0~,
    \label{eqn:p=0cond} \\
    &\mathrm{Re} \left(\sqrt{\det g}\, g^{ij}\right) > 0~. 
     \label{eqn:p=1cond}
\end{align}
The $p=1$ condition is equivalent to requiring that the metric, weighted by the determinant, has positive real part
\begin{equation}\label{p1cond}
    \mathrm{Re} \left(\frac{1}{\sqrt{\det g}}\, g_{ij}\right) > 0~.
\end{equation}
We define the matrix $W$
\begin{equation}\label{wdef}
    W \equiv \frac{1}{\sqrt{\det g}}\, g_{ij} = A + iB~,
\end{equation}
where $A$ and $B$ are the real and imaginary parts of the symmetric matrix $W$. The $p=1$ condition requires strictly positive real part $A > 0$. Since $A$ is real and symmetric it has conventional eigenvalues which, according to the KSW condition, must be positive. 

Once the $p=0$ and $p=1$ conditions are verified, we determine whether \eqref{KSW2} imposes further constraints. To compute the diagonal entries $\lambda_i$, we use the following theorem from linear algebra (Theorem 7.6.4 in \cite{Horn_Johnson_1985}):

\begin{quote}
    Let $A$ and $B$ be real symmetric matrices of the same dimension, with $A$ being positive definite. Then there exists a non-singular matrix $C$ such that
    \begin{equation}\label{thrm}
        C^T AC = I ~, \quad \text{and} \quad C^TBC = K~,
    \end{equation}
    where $K$ is a real diagonal matrix.
\end{quote}
The diagonal entries of $K$ are the generalized eigenvalues $\kappa_i$ of $B$ with respect to $A$, defined as solutions to the generalized eigenvalue problem $B x = \kappa A x$ where the generalized eigenvector $x \neq 0$. They are determined by the characteristic equation
\begin{equation}
    \det(B - \kappa A) = 0~.
\end{equation}
Applying this theorem to $W$, we find that in the diagonal basis
\begin{equation}
    W = (1+i\kappa_i) \delta_{ij}~.
\end{equation}
Consequently, the metric in this diagonal basis takes the form:
\begin{equation}
    g_{ij} = \sqrt{\det g} (1+i\kappa_i) \delta_{ij} \equiv \lambda_i \delta_{ij}~.
\end{equation}
The KSW admissibility condition \eqref{KSW2} is finally expressed as
\begin{equation}
    \sum_i \left|\arg \sqrt{\det g} + \arg (1+i\kappa_i)\right| < \pi~.
\end{equation}

\section{$AdS_5$ Black Holes with Unequal Angular Momenta}\label{sec:uneqj}

In this section we summarize the properties of the $AdS_5$ black hole solutions with unequal angular momenta, followed by their analytic continuations to the Euclidean signature. We focus on the non-extremal, supersymmetric configurations which dominate the gravitational index. We then apply the KSW conditions to the Euclidean black hole saddles, and show that they violate the criteria precisely in the region of chemical potential space where the superconformal index diverges.

\subsection{Lorentzian Black Hole Solutions}
The $AdS_5$ black hole solutions with unequal angular momenta and a single electric charge are solutions to minimal gauged supergravity in $D=5$ spacetime dimensions. They were first written in \cite{Chong:2005hr} and we reproduce them here:\footnote{We put the length of the $AdS_5$ to $l_{AdS_5} = 1$ and further divide the gauge field of \cite{Chong:2005hr} by $\sqrt3$.} 
\begin{align}
        ds^2 =  &-\frac{\Delta_\theta \left[(1+r^2)\rho^2 dt+2q \nu\right]dt}{(1-a^2)(1-b^2)\rho^2}+ \frac{2\,q\,\nu\,\omega}{\rho^2} + \frac{f}{\rho^4}\left(\frac{\Delta_\theta dt}{(1-a^2)(1-b^2)}-\omega\right)^2 \nonumber\\
    &+ \frac{r^2+a^2}{1-a^2}\sin^2\theta d\phi^2 + \frac{r^2+b^2}{1-b^2}\cos^2\theta d\psi^2 +\rho^2 \left(\frac{dr^2}{\Delta_r} + \frac{d\theta^2}{\Delta_\theta}\right) \label{lormet}~,\\
    A&= \frac{q}{\rho^2}\left(\frac{\Delta_\theta dt}{(1-a^2)(1-b^2)}-\omega\right)~,
\end{align}
where
\begin{align}
    \nu &= b\sin^2\theta d\phi +a\cos^2\theta d\psi~,\\
    \omega &= \frac{a\sin^2\theta}{1-a^2} d\phi+ \frac{b\cos^2\theta}{1-b^2} d\psi~,\\
    \Delta_\theta &= 1-a^2\cos^2\theta -b^2\sin^2\theta ~,\\
    \Delta_r &= \frac{(r^2+a^2)(r^2+b^2)(1+r^2)+q^2+2abq}{r^2} -2m~,\label{Deltar}\\
    \rho^2&= r^2 + a^2\cos^2\theta +b^2\sin^2\theta~,\\
    f&= 2m \rho^2 -q^2 + 2abq\rho^2~.
\end{align}
The solutions are expressed in terms of four parameters $(m,a,b,q)$. The four conserved charges $(E,J_a,J_b,Q)$ depend on these four parameters as\footnote{The 5D Newton's constant is put to $G_N=1$ in this paper.}
\begin{align}
    E &= \frac{\pi}{4} \frac{m(3-a^2-b^2-a^2b^2)+2qab(2-a^2-b^2)}{(1-a^2)^2(1-b^2)^2}~,\\
    J_a &= \frac{\pi}{4} \frac{2am+qb(1+a^2)}{(1-a^2)^2(1-b^2)}~,\\
    J_b &= \frac{\pi}{4} \frac{2bm+qa(1+b^2)}{(1-a^2)(1-b^2)^2}~,\\
    Q &= \frac{\pi}{4} \frac{q}{(1-a^2)(1-b^2)}~.
\end{align}
It is often advantageous to trade the parameter $m$ for $r_+$, the largest root of the horizon equation $\Delta_r = 0$. The definition \eqref{Deltar} of $\Delta_r$ gives: 
\begin{equation}
\label{eqn:mandrplus}
    \Delta_r(r_+) = 0 ~\implies~ m =\frac{(r_+^2+a^2)(r_+^2+b^2)(1+r_+^2)+q^2+2abq}{2r_+^2}~.
\end{equation}
The four thermodynamic potentials conjugate to the four conserved charges are expressed in terms of the parameters $(a,b,r_+,q)$ as \cite{Chen:2005zj} 
\begin{align}
    \beta&= 2\pi \frac{r_+\Big((r_+^2+a^2)(r_+^2+b^2)+abq\Big)}{r_+^4(1+2r_+^2+a^2+b^2)-(ab+q)^2}~,\label{BetaLor}\\
    \Omega_a &= \frac{a(r_+^2+b^2)(1+r_+^2)+bq}{(r_+^2+a^2)(r_+^2+b^2)+abq}~,\label{OmaLor}\\
    \Omega_b &= \frac{b(r_+^2+a^2)(1+r_+^2)+aq}{(r_+^2+a^2)(r_+^2+b^2)+abq}~,\label{OmbLor}\\
    \Phi &= \frac{q r_+^2}{(r_+^2+a^2)(r_+^2+b^2)+abq}\label{PhiLor}~.
\end{align}
The entropy of the black holes is given by
\begin{align}
    S = \frac{1}{4} \pi^2\frac{ (r_+^2+a^2)(r_+^2+b^2)+abq}{(1-a^2)(1-b^2)r_+}~.
\end{align}
The black hole solutions are presented in a frame that is static at the asymptotic boundary and with electric potential that vanishes asymptotically. These are the conventional gauge choices.

Supersymmetry can be imposed by demanding that the conserved charges obey the BPS condition
\begin{equation}
\label{BPScondn}
    E = J_a+J_b+3Q~.
\end{equation}
It is equivalent to the following relation between the parameters
\begin{equation}\label{susycondn}
    m = q(1+a+b)~.
\end{equation}
Since $m$ is also given in \eqref{eqn:mandrplus} we can eliminate $r_+$, and describe the supersymmetric solutions by the three parameters $(a,b,q)$.

Before imposing supersymmetry, we can rewrite the conformal factor \eqref{deltar} as: 
\begin{equation}\label{drrewrite}
    \Delta_r = (r^2-r_*^2)^2 + \frac{1}{r^2}\Big(q-q_* -(r^2-r_*^2)(1+a+b)\Big)^2-2\left(m-q(1+a+b)\right)~,
\end{equation}
where
\begin{align}
    r_*^2 &= a+b+ab~,\label{rstardef}\\
    q_* &= (1+a)(1+b)(a+b)~.\label{qstardef}
\end{align}
Supersymmetric solutions satisfy \eqref{susycondn} so the third term in \eqref{drrewrite} vanishes. Then the function $\Delta_r$ reduces to a sum of two squares
\begin{equation}\label{drsusy}
    \Delta_r^{*} = (r^2-r_*^2)^2 + \frac{1}{r^2}\Big(q-q_* -(r^2-r_*^2)(1+a+b)\Big)^2~.
\end{equation}
The conformal factor $\Delta_r$ vanishes at the root $r=r_+$ so, for real Lorentzian solutions, both of the squares in \eqref{drsusy} must vanish at $r=r_+$. Thus the single equation imposed by supersymmetry \eqref{susycondn} imposes {\it two} conditions
\begin{align}
    r_+ &= r_* ~,\label{susycondn1}\\
    q &= q_* ~,
\end{align}
where $r_*$ and $q_*$ refer to the abbreviations in (\ref{rstardef}-\ref{qstardef}). 
The real Lorentzian supersymmetric solutions therefore depend on only two parameters $(a,b)$ and can be obtained from \eqref{lormet} by eliminating the other parameters $(m,q)$ in terms of $(a,b)$ using
\begin{align}
    m &= (1+a)(1+b)(a+b)(1+a+b)~,\\
    q &= (1+a)(1+b)(a+b)~.
\end{align}
For the first square in \eqref{drsusy} to vanish, $r_*^2$ must be non-negative. 
Therefore, the condition that the event horizon at $r=r_+$ is in the real domain, imposes the condition $a+b+ab\geq 0$. 

We often need the parameter $r_+$ which, for the real Lorentzian supersymmetric solutions, is given in terms of $(a,b)$  through \eqref{susycondn1} and so
\begin{equation}
    r_+ = \sqrt{a+b+ab}~.
\end{equation}
With this value, $r=r_+=r_*$ becomes a double root of $\Delta_r^* =0$. Therefore, the black hole is extremal. Indeed, the inverse temperature $\beta$ \eqref{BetaLor} diverges, as expected for an extremal black hole. The thermodynamic potentials (\ref{OmaLor}-\ref{PhiLor}) also simplify and take the values $\Omega^*_a=\Omega^*_b=\Phi^*=1$.

\subsection{Euclidean Saddle Points for the Index}\label{sec:euclsad}

In this subsection we analytically continue the real Lorentzian black holes from the previous subsection in two ways: we allow some parameters in the solutions to be complex, and we continue the black holes to Euclidean signature. The construction will preserve supersymmetry, but the extremality condition $\beta=\infty$ is relaxed. The specific continuation we discuss extends the family of Lorentzian solutions parametrized by two real variables to a family of Euclidean saddle points that depends on three real parameters. 

As before, we do not need to consider spinors explicitly, it is sufficient to impose the BPS condition \eqref{BPScondn} on the conserved charges. We do so by eliminating the parameter $m$ through \eqref{susycondn}, which leaves three parameters $(a,b,q)$. We keep $a$ and $b$ real, so $r^2_*$ and $q_*$ defined in (\ref{rstardef}-\ref{qstardef}) are real as well. We also keep the coordinate $r$ real, and so the position of the horizon $r_+$ must be real. The important relaxation is to permit complex $q$. Then the horizon equation $\Delta_r^*(r_+)=0$ where $\Delta_r^*$ is the sum of two squares \eqref{drsusy} gives a single condition
\begin{equation}
    q = q_* + (r_+^2-r_*^2)(1+a+b \pm i r_+) = -(a\pm ir_+)(b\pm ir_+)(1\pm i r_+)~,
    \label{qsusyres}
\end{equation}
rather than two. This procedure identifies a supersymmetric family of complex solutions that depends on three the real parameters $(a,b,r_+)$. They are restricted to the region defined by the inequalities
\begin{equation}\label{paramineq}
    a^2<1 \qquad b^2<1 \qquad  a b + a +b>0~,
\end{equation}
which are equivalent to: 
\begin{equation}
    - \frac12<b<1, \qquad -\frac{b}{1+b}<a<1~.
    \label{eqn:ineqs}
\end{equation}
Substituting \eqref{qsusyres} back in the definition of $\Delta_r^{*}$ we get
\begin{equation}
\label{deltarEuc}
    \Delta_r^{*} = (r^2-r_*^2)^2 + \frac{1}{r^2}\Big((r_+^2-r^2)(1+a+b)\pm ir_+ (r_+^2-r_*^2)\Big)^2~.
\end{equation}

The $r_+$ and $r_*$ are both real, but generally $r_+\neq r_*$. The real supersymmetric black holes correspond to the special case $r_+=r_*$. It can be convenient to work with the parameters $(a,r_*,r_+)$ by using the definition of $r_*$ in \eqref{rstardef} to eliminate the parameter $b$ as
\begin{equation} \label{belim}
    b = \frac{r_*^2-a}{1+a}~.
\end{equation}
In the new parameterization the inequalities \eqref{eqn:ineqs} can be written as
\begin{align}
    0<r_*<\sqrt3~, \qquad \frac{r_*^2-1}{2}<a<1~.\label{plotineq}
\end{align}
We will write our formulae with both $b$ and $r_*$ in them. Depending on the convention, one can either use \eqref{rstardef} or \eqref{belim} to eliminate one of these two parameters, as needed.

After continuation to complex parameters, the thermodynamics potentials defined 
in (\ref{BetaLor}-\ref{PhiLor}) become: 
\begin{align}
    \beta&= 2\pi \frac{(a\pm ir_+)(b\pm ir_+)(r_*^2\mp ir_+)}{(r_+^2-r_*^2)\left[2(1+a+b)r_+ \mp i(r_*^2-3r_+^2)\right]}~,
    \label{betaEuc}
    \\
    \Omega_a &= \frac{(r_*^2\mp i a r_+)(1\pm ir_+)}{(r_*^2\mp ir_+)(a\pm ir_+)}~,\label{OmegaaEuc}\\
    \Omega_b &=\frac{(r_*^2\mp i b r_+)(1\pm ir_+)}{(r_*^2\mp ir_+)(b\pm ir_+)}~,\label{OmegabEuc}\\
    \Phi &= \frac{r_+(r_+\mp i)}{r_*^2\mp ir_+ }~.
\end{align}
They satisfy the constraint
\begin{equation}
    \beta(1+\Omega_a+\Omega_b - 3\Phi) = \mp 2\pi i~,
\end{equation}
as required by the boundary conditions of the Euclidean Gravitational Path Integral computing the superconformal index.

The metric for the saddle point of the index can be obtained from the Lorentzian metric for the supersymmetric black holes given in \eqref{lormet} in two steps. First recall that the Lorentizian metric is given in a coordinate system where it becomes static at infinity i.e. asymptotically the metric becomes that of $AdS_5$. In such a frame, there is an ergoregion around the horizon. The replacements $\phi \rightarrow \phi +\Omega_a t$ and $\psi \rightarrow \psi + \Omega_bt$ with $t$ fixed define a co-rotating coordinate system so that the time coordinate $t$ is timelike everywhere outside the horizon. Then the analytic continuation $t\rightarrow -i\beta t_E$ can be performed to obtain the complex Euclidean solutions with metric
\begin{align}
    ds_E^2 =  &\frac{\Delta_\theta \beta\left[(1+r^2)\rho^2\beta dt_E+2iq \nu\right]dt_E}{(1-a^2)(1-b^2)\rho^2}+ \frac{2\,q\,\nu\,\omega}{\rho^2} + \frac{f}{\rho^4}\left(\frac{i \beta \Delta_\theta dt_E}{(1-a^2)(1-b^2)}+\omega\right)^2 \nonumber\\
    &+ \frac{r^2+a^2}{1-a^2}\sin^2\theta \left(d\phi-i\Omega_a\beta dt_E\right)^2 + \frac{r^2+b^2}{1-b^2}\cos^2\theta \left(d\psi-i\beta\Omega_bdt_E\right)^2 \nonumber\\&+\rho^2 \left(\frac{dr^2}{\Delta_r} + \frac{d\theta^2}{\Delta_\theta}\right) \label{metricuneqj}~,
\end{align}
where
\begin{align}
    \nu &= b\sin^2\theta \left(d\phi - i\beta\Omega_a dt_E\right)+a\cos^2\theta \left(d\psi - i\beta\Omega_b dt_E\right)~,\\
    \omega &= \frac{a\sin^2\theta}{1-a^2} \left(d\phi - i\beta\Omega_a dt_E\right)+ \frac{b\cos^2\theta}{1-b^2} \left(d\psi - i\beta\Omega_b dt_E\right)~,\\
    \Delta_\theta &= 1-a^2\cos^2\theta -b^2\sin^2\theta ~,\\
    \Delta_r &= \frac{(r^2+a^2)(r^2+b^2)(1+r^2)+q^2+2abq}{r^2} -2m \label{deltar}~,\\
    \rho^2&= r^2 + a^2\cos^2\theta +b^2\sin^2\theta~,\\
    f&= 2m \rho^2 -q^2 + 2abq\rho^2~,
\end{align}
and $(m,a,b,q)$ are parameters in the solution and the quantities $(\beta,\Omega_a,\Omega_b)$ are defined in \eqref{betaEuc}-\eqref{OmegabEuc}. The ranges of the coordinates are: $r\in \mathbb R^+,\, t_E \sim t_E+1,\, \theta\in\left[0,\frac\pi2\right], \, \phi,\psi\in \left[0,2\pi\right]$.

The metric \eqref{metricuneqj} has simple subdeterminants: 
\begin{align}
    \det g_{(t,\phi,\psi)} &= \beta^2 \frac{\Delta_r\Delta_\theta r^2 \sin^2 2\theta}{4(1-a^2)^2(1-b^2)^2}~,
    \label{detphipsi}\\
    \det g_{(r,\theta)} &= \frac{\rho^4}{\Delta_r\Delta_\theta}~.
    \label{detrtheta}
\end{align}
The full determinant becomes: 
\begin{equation}\label{detdef}
   \det g = \frac{\beta^2 \rho^4 r^2 \sin^2 2\theta}{4(1-a^2)^2(1-b^2)^2}~.
\end{equation}

\subsection{Stability of the Euclidean Saddle Points for the Index}
\label{sec:asympcond}
In this section we apply the KSW criterion detailed in Section \ref{sec:KSW} to the complex Euclidean saddle points of the gravitational index described in Section \ref{sec:euclsad}. We show that it is both necessary and sufficient to identify the complex saddles that lie in the region where the index is well defined. The show that the instabilities first arise far from the black hole core, in the plane the dominant angular momentum. 

\subsubsection{$p=0$ KSW condition}

The $p=0$ KSW condition \eqref{eqn:p=0cond} reads
\begin{equation}
    {\rm Re}\, \left(\sqrt{\det g}\right) > 0 ~.
\end{equation}
Since $\rho^2>0$, $r>0$ and $\theta\in \left[0,\frac\pi2\right]$, the square root of the expression for the determinant \eqref{detdef} gives
\begin{equation}
    \sqrt{\det g} = \frac{\beta \rho^2 r \sin 2\theta}{2(1-a^2)(1-b^2)}~.
\end{equation}
We picked the branch of the square root that is inherited from Lorentzian solutions which must have non-negative temperature. We find that the $p=0$ KSW condition reduces to
\begin{equation}
    {\rm Re}\,(\beta) >0~.
    \label{eqn:rebetapos}
\end{equation}
This is a requirement on the Euclidean gravitational path integral that mirrors the restriction (\ref{betabound}) needed for a well-defined superconformal index. 

The explicit expression for the real part of the inverse temperature $\beta$ is: 
\begin{equation}
    {\rm Re} (\beta) = \frac{2 \pi  r_+ \left(-a^3 \left(1-b^2\right)+\left(a^2 (b+3) \left(b^2+r_+^2\right)\right)-a \left(1-b^2\right) r_+^2+\left(b^2+r_+^2\right) \left(3 r_+^2-b\right)\right)}{\left(r_+^2- a b-a-b\right) \left(2 r_+^2 \left(2 a^2+a b+a+2 b^2+b+2\right)+(a b+a+b)^2+9 r_+^4\right)}~.
    \label{eqn:fullrebeta}
\end{equation}
The condition \eqref{eqn:rebetapos} is satisfied if we add the constraint
\begin{equation}\label{p0const}
    r_+ >r_* =\sqrt{a b+a + b} ~,
\end{equation}
to the bounds \eqref{paramineq} on the parameters $(a,b,r_+)$. 
Thus the three parameter family of Euclidean solutions have a coordinate position $r_+$ of the horizon that has moved {\it outwards} from the location $r_*$ that is required by the supersymmetry condition \eqref{susycondn1} on the two parameter family of Lorentzian black holes.

\subsubsection{$p=1$ KSW condition}
The $p=1$ KSW condition reads
\begin{equation}
    {\rm Re}\,\left(\frac{1}{\sqrt{\det g}}g_{ij}\right)>0~.
\end{equation}
The Euclidean metric $g$ has the form
\begin{equation}
    g = \begin{pmatrix}
        -\beta^2 g_{tt} & -i\beta g_{t\phi} &-i\beta g_{t\psi} & 0&0\\
        -i\beta g_{t\phi} & g_{\phi\phi} & g_{\phi\psi} & 0&0\\
        -i\beta g_{t\psi} & g_{\phi\psi} & g_{\psi\psi} & 0&0\\
        0&0&0 &g_{rr} &0\\
        0&0&0  &0&g_{\theta\theta}
    \end{pmatrix}~,
\end{equation}
where the quantities $g_{tt},g_{t\phi},\ldots$ refer to the components of the Lorentzian metric after complexification to a family supersymmetric non-extremal black holes, but before analytical continuation to Euclidean signature.

The simplest components of the $p=1$ conditions are the $\theta$ and $r$ directions. First we have 
\begin{align}
    {\rm Re} \, \left(\frac{g_{\theta\theta}}{\sqrt{\det g}}\right)>0 ~& \implies  {\rm Re} \, \left(\beta\right)>0~, 
\end{align}
which is simply the $p=0$ condition \eqref{eqn:rebetapos}. Next, we find: 
\begin{align}
    {\rm Re} \, \left(\frac{g_{rr}}{\sqrt{\det g}}\right)>0 ~& \implies ~{\rm Re} \, \left(\beta\Delta_r^{*}\right)>0~.
\end{align}
This is more complicated. Taking $\beta$ from \eqref{betaEuc} and $\Delta_r^*$ from \eqref{deltarEuc}, it is straightforward to compute ${\rm Re} \, (\beta\Delta_r^*)$. The resulting expression is unwieldy, so we do not present it here. The significant feature is that it is proportional to $(r_+-r_*)^{-1}$, with an overall factor that is positive at $r=r_+$ and a monotonically increasing function of $r$. After \eqref{eqn:fullrebeta} we concluded that $r_+-r_*$ and ${\rm Re} \, (\beta)$ have the same sign, and here find that ${\rm Re} \, (\beta\Delta_r^*)$ has the very same sign. Therefore, the $r$ component of the $p=1$ KSW condition reduces to the $p=0$ KSW condition \eqref{eqn:rebetapos} once again.

The metric along the $(t,\phi,\psi)$ directions is too complicated to analyze analytically at a general point in the geometry. Therefore, in the remainder of this subsection, we focus on the asymptotic region near the boundary of $AdS$. In subsection \ref{sec:bulkcond} we argue that this gives the dominant condition on the entire spacetime. 

In the asymptotic region $r\rightarrow\infty$, the metric \eqref{metricuneqj} can be written in the form
\begin{equation}
    ds^2_E \sim \frac{1}{z^2}dz^2+\frac{1}{z^2}\Big(\beta^2dt_E^2+d\vartheta^2+\sin^2\vartheta(d\phi-i\beta\Omega_a dt_E)^2+\cos^2\vartheta(d\psi-i\beta\Omega_b dt_E)^2\Big)~,
\end{equation}
where the asymptotic coordinates $(z,\vartheta)$ are defined in terms of $(r,\theta)$ through
\begin{equation}
    \frac{(1-a^2)\sin^2\vartheta}{z^2} = (r^2+a^2)\sin^2\theta~, \qquad \frac{(1-b^2)\cos^2\vartheta}{z^2} = (r^2+b^2)\cos^2\theta~.
\end{equation}
The determinant \eqref{detphipsi} of the metric along $(t_E,\phi,\psi)$ takes the simple form
\begin{equation}
    \sqrt{\det g} \sim  \beta \sin 2\vartheta~,
\end{equation}
up to an overall conformal factor that depends on $z$.
The $p=1$ KSW condition then reads 
\begin{align}
    {\rm Re}\, \begin{pmatrix}
        \beta(1-\sin^2\vartheta\, \Omega_a^2 -\cos^2\vartheta \,\Omega_b^2) & -i \sin^2\vartheta \, \Omega_a & -i \cos^2\vartheta \, \Omega_b\\
         -i \sin^2\vartheta \, \Omega_a & \frac1\beta\sin^2\vartheta & 0 \\
        -i \cos^2\vartheta \, \Omega_b & 0 & \frac1\beta \cos^2\vartheta
    \end{pmatrix} = {\rm Re}\, (A+iB)>0\label{p1kswmat}~. 
\end{align}
We freely omitted an overall conformal factor that is real and positive. 
The real symmetric matrices $A$ and $B$ are introduced through $W=A+iB$, as in \eqref{wdef}. 

The matrix $A$ is an arrowhead matrix: it has non-zero entries only in the first row, the first column, and the diagonal. For such matrices, the positivity condition reduces to: 
\begin{align}
    a_{ii} >&\ 0 \qquad i=2,\ldots,n~,\label{arrow1}\\ 
    a_{11} >& \sum_{j=2}^n \frac{a_{1j}^2}{a_{jj}}~, \label{arrow2}
\end{align}
where $a_{ij}$ ($i,j=1,2,\ldots,n$) are elements of the matrix $A$.

The conditions \eqref{arrow1} on the diagonal elements of the matrix \eqref{p1kswmat} reduces to the condition ${\rm Re}\, \beta>0$ that was imposed repeatedly already. The second condition \eqref{arrow2} will prove non-trivial. After some algebra, it reduces to: 
\begin{equation}
    \sin^2 \vartheta\ {\rm Re}\left[\beta(1-\Omega_a)\right]\ {\rm Re}\left[\beta(1+\Omega_a)\right] + \cos^2 \vartheta\ {\rm Re}\left[\beta(1-\Omega_b)\right]\ {\rm Re}\left[\beta(1+\Omega_b)\right] >0 ~.\label{uneqjksw}
\end{equation}
This equation must apply for all all $\vartheta\in \left[0,\frac\pi2\right]$, so it is equivalent to the two conditions,
\begin{align}
    {\rm Re}\left[\beta(1-\Omega_a)\right]\ {\rm Re}\left[\beta(1+\Omega_a)\right]>0~,\\
    {\rm Re}\left[\beta(1-\Omega_b)\right]\ {\rm Re}\left[\beta(1+\Omega_b)\right]>0~.
\end{align}
In the original Lorentzian solutions we picked positive $\Omega_a$ and $\Omega_b$ without loss of generality, so we intuit that, in each of these equations, it is the first factor that is decisive. We are unable to prove this analytically, but we have checked with Mathematica that the expressions ${\rm Re}\left[\beta(1+\Omega_a)\right]$ and ${\rm Re}\left[\beta(1+\Omega_b)\right]$ are positive throughout the parameter space defined by \eqref{paramineq} and \eqref{p0const}. Therefore, for the complex Euclidean saddle points of the Superconformal Index, the $p=1$ KSW condition in the asymptotic region reduces to
\begin{align}
    {\rm Re}\left[\beta(1-\Omega_a)\right]>0~,\\
    {\rm Re}\left[\beta(1-\Omega_b)\right]>0~.
\end{align}
These two inequalities agree with \eqref{omegabound}, with the identifications \eqref{omegaiden}. In Section \ref{sec:converge}, these bounds were imposed in order for the superconformal index to converge. In other words, the KSW condition for the complex Euclidean saddle points breaks down exactly at the boundary of the region of convergence of the index.

\begin{figure}
    \centering
    \includegraphics[width=0.9\linewidth]{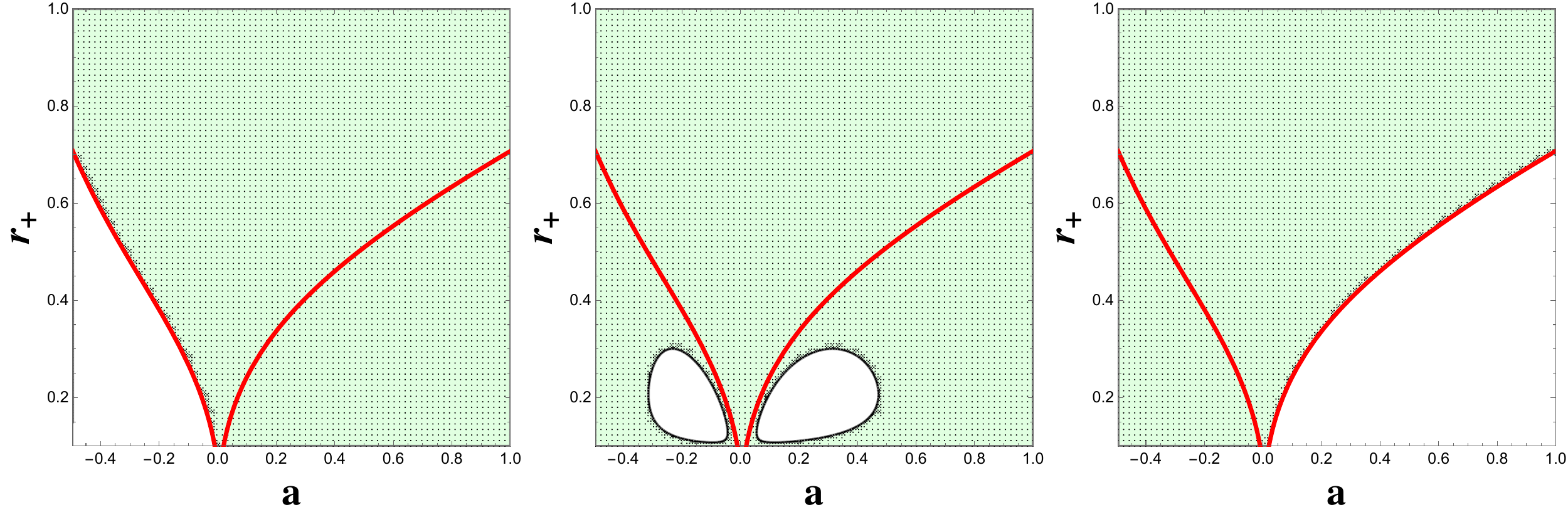}
    \caption{The $(a, r_+)$ cross sections of the parameter space at constant $r_*=0.1$. The green shaded region represents the parameter space where the $p=1$ KSW criterion is satisfied in the asymptotic AdS$_5$ region of the complex metric. The plots from the left to right take $\vartheta=0,\pi/4,\pi/2$. The KSW condition is satisfied for all $\vartheta$ inside the red curves. These are precisely the boundaries of the region of convergence of the superconformal index. The plots were obtained using the function \texttt{RegionPlot} in \texttt{Mathematica} with an initial grid of $2500$ points and with a \texttt{MaxRecursion} of 2.}
    \label{fig:KSWinf}
\end{figure}

We validated our calculations numerically, with results presented in Figure \ref{fig:KSWinf}. The three real parameters $(a,r_+,r_*)$ on configuration space are visualized in plots as the $(a,r_+)$ cross-section at a fixed value of $r_*$. At fixed $r_*$, the range of the parameter $r_+$ is $r_+>r_*$, because of the $p=0$ condition \eqref{p0const}, and the parameter $a$ takes values between $a_{\text{min}}=(r_*^2-1)/2$ and $a_{\text{max}}=1$, as shown in \eqref{plotineq}. We picked the value $r_*=0.1$ in all plots. 
The green regions in the three plots correspond to parameters where \eqref{uneqjksw} holds at three different values of $\vartheta = 0,\frac\pi4,\frac\pi2$. The strongest constraints are obtained at $\vartheta=0$ and $\vartheta=\frac\pi2$. Taking the intersection of both these regions we see that the region of validity of the $p=1$ KSW condition is bounded by the curves colored red. These curves are precisely where the real parts of the renormalized potentials $\beta(1-\Omega_a)$ or $\beta(1-\Omega_b)$ vanish. In other words, the $p=1$ KSW criterion excludes the region where a Hilbert space interpretation of the index breaks down because the real parts of the renormalized chemical potentials $\Omega_{a,b}^\prime$ are negative.

\subsubsection{$p\geq 2$ KSW conditions}
Our analysis of the $p=1$ KSW conditions was explicit, working directly with the Euclidean metric. 
As explained in section \ref{sec:KSW}, we check the $p=2$ KSW conditions using a different strategy, by employing a basis where the metric takes the diagonal form $g_{ij}=\lambda_i\delta_{ij}$. As before, the complexity of the metric at finite radial coordinate motivate us to first perform our analysis in the asymptotic region. In subsection \ref{sec:bulkcond} we argue that this gives the dominant condition on the entire spacetime. 

The metric is diagonal and real in the $(r,\theta)$ directions so $|\arg \lambda_{r,\theta}|=0$. Therefore, we can focus on the $3\times 3$ block of the metric along the directions $(t_E,\phi,\psi)$. The diagonal values $\lambda_{t_E,\phi,\psi}$ of the metric in the diagonal basis are determined through the procedure detailed in section \ref{sec:KSW}.  The matrix $W$ defined in \eqref{wdef} is given by 
\begin{equation}
    W = \begin{pmatrix}
        \beta(1-\sin^2\vartheta\, \Omega_a^2 -\cos^2\vartheta \,\Omega_b^2) & -i \sin^2\vartheta \, \Omega_a & -i \cos^2\vartheta \, \Omega_b\\
         -i \sin^2\vartheta \, \Omega_a & \frac1\beta\sin^2\vartheta & 0 \\
        -i \cos^2\vartheta \, \Omega_b & 0 & \frac1\beta \cos^2\vartheta
    \end{pmatrix} \equiv A+iB~. 
\end{equation}
The symmetric matrices $A$ and $B$ are the real and imaginary parts of $W$. The generalized eigenvalues $\kappa_i$ of $B$ with respect to $A$ are given by the roots of the equation
\begin{equation}
    \det(B-\kappa A) =0~.
    \label{eqn:geneigen}
\end{equation}
We focus on the two 2-planes at $\vartheta=0$ and $\vartheta=\frac\pi2$ where the $p=1$ KSW conditions give the strongest constraints. Then the metric reduces to the $2\times 2$ block
\begin{equation}\label{p2wmat}
    W=\begin{pmatrix}
        \beta(1- \Omega_{a,b}^2) & -i \Omega_{a,b} \\
         -i \Omega_{a,b} & \frac1\beta
    \end{pmatrix} \equiv A+iB~.
\end{equation}
and the generalized eigenvalue equation \eqref{eqn:geneigen} becomes a quadratic equation in $\kappa$.
Substituting the matrices $A,B$ from \eqref{p2wmat}, we find after some algebra that the coefficient of the linear term in $\kappa$ vanishes, and the quadratic equation becomes
\begin{equation}
    \left(\mathrm{Re}\left[\frac{1}{\beta}\right]\mathrm{Re}\left[\beta(1-\Omega_a^2)\right]-(\mathrm{Im}\left[\Omega_a\right])^2\right) \kappa^2 = (\mathrm{Re}\left[\Omega_a\right])^2 - \mathrm{Im}\left[\frac1\beta\right]\mathrm{Im}\left[\beta(1-\Omega_a^2)\right] ~.  
\end{equation}
This equation has two real roots ($\kappa_1,-\kappa_1$) that are equal in magnitude but opposite in sign. The reality of the roots in the region where the $p=1$ KSW condition holds is guaranteed by the result \eqref{thrm} given in section \ref{sec:KSW}. Therefore the $p\geq2$ KSW conditions can be written as
\begin{equation}
    |\arg \beta + \arg(1+i \kappa_1)| + |\arg \beta - \arg(1+i \kappa_1)| < \pi~,
\end{equation}
which further reduce to
\begin{align}
    |\arg \beta|<\frac\pi2 \qquad \text{ and } \qquad |\arg (1+i\kappa_1)|<\frac\pi2~.
\end{align}
The first inequality reduces to the $p=0$ condition \eqref{eqn:rebetapos} and the 
second inequality is satisfied trivially, since $\kappa_1$ is real.

\subsection{KSW Conditions in the Bulk of Spacetime}
\label{sec:bulkcond}
The analytical study of KSW conditions in the previous section was performed near the asymptotic boundary of the geometry. We now give numerical evidence that the most stringent constraints are in fact obtained in the asymptotic region. 

The practical procedure outlined in section \ref{sec:KSW} is sequential: we begin by checking the $p=0$ condition and proceed to check the $p=1$ criterion only if the $p=0$ condition is satisfied, and then finally check the $p\geq 2$ conditions only if both the $p=0$ and $p=1$ conditions are satisfied. 

\begin{figure}
    \centering
    \includegraphics[width=0.9\linewidth]{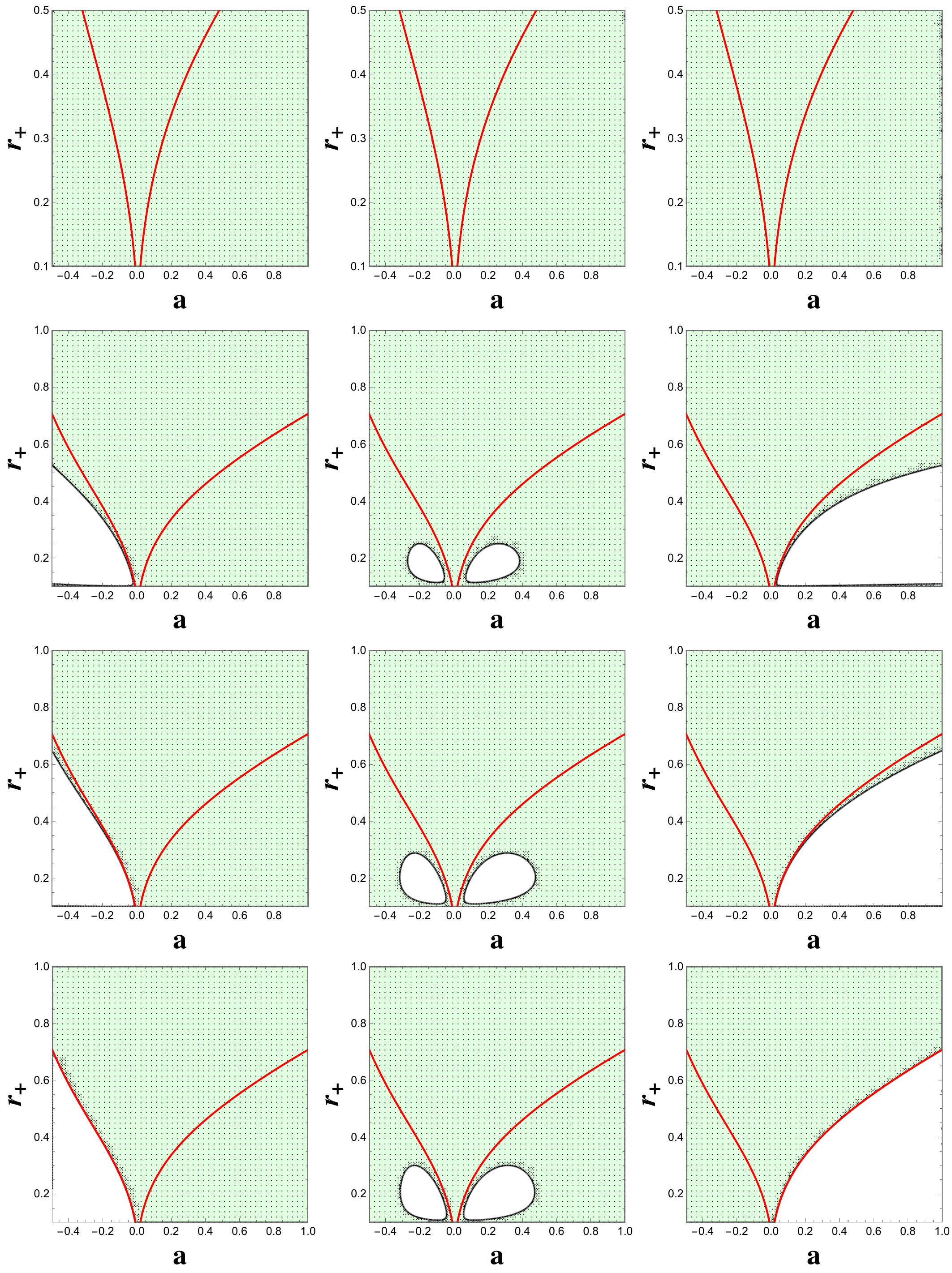}
    \caption{The plots show $(a, r_+)$ cross sections of the parameter space at constant $r_*=0.1$. The green region represents the region of parameter space where the corresponding black hole saddle point satisfies all $p\geq0$ KSW conditions. The red curves are the boundaries of the region of convergence of the index. The columns from the left to right show the validity of the KSW condition at $\vartheta=0,\pi/4,\pi/2$. The rows from top to bottom show the validity of the KSW condition at $r=0.5,2,4,\infty$. We observe that the strongest constraints are obtained at $r=\infty$ and $\vartheta = 0,\pi/2$. The plots were obtained using the function \texttt{RegionPlot} in \texttt{Mathematica} with an initial grid of $2500$ points and with a \texttt{MaxRecursion} of $2$.}
    \label{fig:KSWuneqj}
\end{figure}

The results of this procedure are presented in Figure \ref{fig:KSWuneqj}. The three real parameters $(a,r_+,r_*)$ on configuration space are visualized in plots as the $(a,r_+)$ cross-section at a fixed value of $r_*$. At fixed $r_*$, the range of the parameter $r_+$ is $r_+>r_*$, because of the $p=0$ condition \eqref{p0const}, and the parameter $a$ takes values between $a_{\text{min}}=(r_*^2-1)/2$ and $a_{\text{max}}=1$, as shown in \eqref{plotineq}. We picked the value $r_*=0.1$ in all plots. 

The KSW conditions \eqref{KSW1} must be satisfied pointwise in the complex Euclidean spacetime so, for each value of the three real parameters $(a,r_+,r_*)$, we should consider all values of 
the angular coordinate $\vartheta$, and the entire range $r>r_+$ of the radial coordinate $r$, with the lower bound so it describes a point outside the event horizon. When reading Figure \ref{fig:KSWuneqj} from top to bottom, the value of the radial coordinate for each row of plots is fixed at $r=0.5,2,4,\infty$. In each row, the plots are at fixed value of $\vartheta = 0, \pi/4, \pi/2$. 

In the plots collected in Figure \ref{fig:KSWuneqj}, the regions where all the KSW conditions are satisfied are colored green. In our computations we found that the $p=0,1$ conditions are exhaustive, the $p\geq 2$ conditions do not add anything new. This generalizes the analytical results near the asymptotic boundary presented in the previous section to the bulk of the geometry. In Figure \ref{fig:KSWuneqj}, we observe the following features: 
\begin{itemize}
    \item The KSW conditions become ever more stringent as one reads the Figure from top to bottom. This corresponds to motion from the horizon and the asymptotic boundary. The strongest conditions are obtained at the asymptotic boundary, shown in the last row of plots.
    \item In the angular coordinate the strongest conditions are obtained at $\vartheta = 0,\pi/2$. Those are the first and third columns of the plots.
    \item The strongest conditions show that the KSW criteria are satisfied everywhere in the geometry precisely when the parameters are inside the red curves. These are the boundaries of the region of convergence of the index, so they indicate the breakdown of the Hilbert space interpretation in the dual CFT. 
\end{itemize}
We conclude that the strongest conditions arise near the asymptotic boundary $r=\infty$, and at the poles $\vartheta=0, \pi/2$. The analytical study in Section \ref{sec:asympcond} focused on $r=\infty$ in anticipation of this result. Accordingly, Figure \ref{fig:KSWinf} is identical to the last row in Figure \ref{fig:KSWuneqj}.

The spatial dependence of the instability is significant. That the breakdown is at large $r$ suggests that the black hole in Lorentzian signature becomes unstable in the region near the asymptotic boundary. That it is in a $2$-planes at $\vartheta = 0,\pi/2$ indicates that the instability is driven by one component of the angular momentum. These features are consistent with a phase transition between the single center black hole and a two component configuration comprising both a black hole core and a BPS gas near the asymptotic boundary that carries large angular momentum efficiently. Such a two component configuration in Lorentzian spacetime is known as a Grey Galaxy\cite{Kim:2023sig,Bajaj:2024utv}.

\section{Discussion}
\label{sec:discussion}
In this work, we have systematically analyzed the complex Euclidean saddle points of the gravitational index, specifically focusing on solutions in $AdS_5$ with unequal angular momenta. Our primary result is a validation of the Kontsevich-Segal-Witten (KSW) criterion: the complex saddles that violate the KSW conditions \eqref{KSW1} lie exactly in the region of the complex chemical potential space where the dual superconformal index is expected to diverge. 

The divergence of the partition function (or index) in the grand canonical ensemble typically signals a breakdown of the saddle point approximation and an instability towards a new phase. Our results suggest that the violation of the KSW conditions is the bulk geometric realization of this instability.
Recently, the microcanonical version of the index has been explored in detail \cite{Choi:2025lck,Deddo:2025jrg,Jones:2025gno}, leading to the identification of novel two-component phases \cite{Kim:2023sig,Choi:2024xnv,Bajaj:2024utv}. These phases typically consist of a black hole at the center of the spacetime surrounded by a gas of gravitons (or a few dual giant gravitons). Notably, these multi-component phases dominate the ensemble in exactly the regions of parameter space where we found that the KSW criteria instruct us to exclude the standard complex Euclidean black hole saddles.
This points to a compelling physical picture:
\begin{itemize}
    \item The standard BPS black hole saddle has a divergent contribution to the path integral (diagnosed by KSW) precisely when the system undergoes a phase transition.
    \item Although these new two-component phases exist in a parameter space where the original saddle was unstable, the new two component configurations that describe the endpoint of this instability must have chemical potentials inside the convergent region of the index.
    \item These new phases were earlier constructed in the Lorentzian signature, where the black hole component is extremal, supersymmetric, and real \cite{Choi:2025lck} .
\end{itemize}
A fully Euclidean understanding of the novel two-component phases remains an open problem. The failure of the KSW criterion for the single-center black hole suggests a new, stable Euclidean saddle — perhaps a complex metric describing a ``black hole plus gas" configuration — that reproduces the features of the microcanonical index.

Recently, \cite{Anand:2025mfh} (building on \cite{Bhattacharyya:2007vs,Banerjee:2012iz,Jensen:2012jh,Shaghoulian:2015lcn,Benjamin:2023qsc}) studied thermal partition functions near the boundaries of the convergence region. For 4D CFTs, they observed a semi-universal behavior as the chemical potentials $\Omega_a^\prime \to 0$ that  is consistent with phase transitions between Kerr-AdS black holes, Grey Galaxies and the pure Thermal AdS gas.
It would be highly desirable to see an explicit phase transition in the superconformal index in the canonical ensemble that mirrors the Grey Galaxy phase transitions seen in the thermal partition function. The numerical evidence from the microcanonical ensemble \cite{Choi:2025lck,Deddo:2025jrg} strongly supports the existence of supersymmetric versions of the Grey Galaxies, but Euclidean saddles describing them are still lacking.

Finally, while this work focused on the KSW criterion, other proposals for determining the admissibility of complex metrics in the gravitational path integral have been explored recently \cite{DiTucci:2020weq,Jonas:2022uqb,Mahajan:2025bzo,Singhi:2025rfy,Ailiga:2025osa}. It would be instructive to apply these alternative diagnostics to the unequal angular momenta saddles studied here. Specifically, if those results can reproduce the same forbidden regions as the KSW criterion, it would provide progress towards a universal rule governing admissibility of complex metrics in Euclidean quantum gravity.

\section*{Acknowledgements}

We thank
D. Cassani,
S. Minwalla,
E. Lee,
L. Pando-Zayas,
and C. Patel
for discussions.
This work was supported in part by DoE grant DE-SC0007859, the Leinweber Institute for Theoretical Physics, and the Department of Physics at the University of Michigan. 

\nocite{*}
\bibliographystyle{JHEP}
\bibliography{refs.bib}

\end{document}